\documentclass{PoS}
\usepackage{amssymb}
\usepackage{amsmath}

\newcommand{\be}{\begin{equation}}
\newcommand{\ee}{\end{equation}}
\newcommand{\bel}[1]{\be\label{#1}}
\newcommand{\re}[1]{Eq.~(\ref{#1})}

\newcommand{\hsp}{\hspace*{1pt}}
\newcommand{\hspm}{\hspace*{.5pt}}
\newcommand{\ov}[1]{\overline{#1}}

\usepackage{graphicx}
\usepackage{bm}

\title{Hydrodynamic modeling of a pure-glue initial scenario in high-energy hadron and heavy-ion collisions}

\ShortTitle{Hydrodynamic modeling of a pure-glue initial scenario}

\author{V.~Vovchenko$^{\,a,b,c}$, Long-Gang~Pang$^{\,a}$, H.~Niemi$^{\,b}$, Iu.A.~Karpenko$^{\,d,e}$, M.I.~Gorenstein$^{\,a,d}$,
L.M.~Satarov$^{\,a,f}$, I.N.~Mishustin$^{\,a,f}$, B.~K\"ampfer$^{\,g,h}$,
\speaker{H.~Stoecker}$^{\,a,b,i}$\\

$^a$\mbox{Frankfurt Institute for Advanced Studies, D-60438 Frankfurt, Germany}\\
$^b$\mbox{Institut f\"ur Theoretische Physik, Goethe Universit\"at, D-60438 Frankfurt, Germany}\\
$^c$\mbox{Department of Physics, Taras Shevchenko National University of Kiev, 03022 Kiev, Ukraine}\\
$^d$\mbox{Bogolyubov Institute for Theoretical Physics, 03680 Kiev, Ukraine}\\
$^e$\mbox{INFN - Sezione di Firenze, I-50019 Sesto Fiorentino (Firenze), Italy}\\
$^f$\mbox{National Research Center ''Kurchatov Institute'', 123182 Moscow, Russia}\\
$^g$\mbox{Helmholtz-Zentrum Dresden-Rossendorf, D-01314 Dresden, Germany}\\
$^h$\mbox{Technische Universit\"at Dresden, Institut f\"ur Theoretische Physik,
D-01062 Dresden, Germany}
$^i$\mbox{GSI Helmholtzzentrum f\"ur Schwerionenforschung GmbH, D-64291 Darmstadt, Germany}
\\
E-mail: \email{H.Stoecker@gsi.de}}

\abstract{
Partonic matter produced in the early stage of ultrarelativistic
nucleus-nucleus collisions
is assumed to be composed mainly of gluons, and quarks and antiquarks are produced at later times.
The comparable hydrodynamic simulations of heavy-ion collisions
for (2+1)-flavor and Yang-Mills equations of state performed by using three different hydrodynamic codes are presented.
Assuming slow chemical equilibration of quarks,
the spectra and elliptic flows of thermal
dileptons and photons are calculated for central Pb+Pb collisions at the LHC energy of
$\sqrt{s_{_{\rm NN}}} = 2.76$~TeV.
It is shown that a suppression of quarks at early times leads to a
significant reduction of the yield of the thermal dileptons, but only to a rather modest suppression
of the $p_T$-distribution of direct photons. It is demonstrated that an
enhancement of photon and dilepton elliptic flows might serve as a promising signature of the pure-glue initial state.
Calculations based on Bjorken hydrodynamics suggest that collisions of small systems at intermediate energies available at RHIC or future FAIR facilities may show stronger effects associated with initial pure gluodynamic evolution.
}

\FullConference{54th International Winter Meeting on Nuclear Physics\\
		25-29 January 2016 \\
		Bormio, Italy}

\begin{document}

\section{Introduction}

There are many different approaches to describe the initial stage of nucleus-nucleus (A+A) collisions.
Usually it is assumed that strong nonequilibrium effects take place
only during a very short proper time interval $\tau_s \sim 1/Q_s$,
where $Q_s\simeq 1\div 2~\textrm{GeV}$ is the so-called gluon saturation
scale~\cite{Gribov}. The idea that the gluonic components of
colliding nucleons dominate in high energy collisions was originally put forward
in Ref.~\cite{Pok-Hove}, and it was motivated
by the fact that the perturbative gluon-gluon cross sections are larger than the quark-antiquark
ones.  A two-step equilibration of QGP was proposed in~\cite{raha,shuryak,sinha} assuming that
the gluon thermalization is accomplished already at the early proper time $\sim$ $\tau_s$, while the quark-antiquark chemical
equilibration proceeds until later times $\tau_{\rm th}>\tau_s$ (according to Ref.~\cite{Xu05},
$\tau_{\rm th}=5\div10~\textrm{fm}/c$).
Possible signatures of quark undersaturation in high energy A+A collisions were considered by different authors, see,
e.g., Refs.~\cite{shuryak,Bir93,Str94,BK1,BK2,Tra96,Ell00,Dut02,Gel04,Liu14,Mon14,Moreau2015}.

Recently, the {\it pure glue} initial scenario of
Pb+Pb collisions at Large Hadron Collider (LHC)
energies was discussed in~Refs.~\cite{Sto16,Sto15}.
Below we describe how a pure-glue initial scenario for heavy-ion collisions can be modeled using the (2+1)--dimensional boost-invariant hydrodynamics. 
In particular, we demonstrate the difference between hydrodynamic evolution of the full QCD matter and of the pure glue matter by performing simulations within three different hydrodynamic codes.
We also introduce
the time-dependent quark-antiquark fugacity in order to describe the QGP
evolution in the absence of the full chemical equilibrium.
The new aspects of the present study include constructing
the equation of state for chemically nonequilibrated QCD matter
via an interpolation of the lattice data, and the solution of hydro equations with a time-dependent equation of state.

A significant emphasis is
put on electromagnetic probes (thermal photons and dileptons), which may carry an important information
about the deconfined phase. This problem has been repeatedly addressed in the literature, see,
e.g., \cite{BK1,BK2,Gel04,Liu14,Mon14}, however, a definitive conclusion about the role of chemically nonequilibrium evolution is still not obtained. 
We also analyze
the impact of chemical nonequilibrium effects on the 
momentum anisotropy of photons and
dileptons.
A more detailed version of this study
is given in Ref.~\cite{Vov16a}.

In addition, we simulate the evolution of system created in A+A and p+p collisions at lower energies, by employing the simplified Bjorken hydrodynamics. There we consider only two limiting cases: the fully equilibrated (2+1)-flavor evolution and the pure gluodynamic evolution.

\section{\protect Formulation of the model}

The equations of the longitudinally boost-invariant (2+1)--dimensional ideal hydrodynamics can be written as (\mbox{$\hbar=c=1$})
\bel{hydro}
\frac{\partial\hsp T^{\mu\nu}}{\partial\hsp x^{\nu}}~=~0~,
\ee
where
$T^{\mu\nu}~=~(\varepsilon + P)\hsp u^{\mu}u^{\nu}~-~Pg^{\mu\nu}~$
is the energy-momentum tensor, $u^\mu$ is the 4-velocity, $\varepsilon$ and $P$ are the local rest-frame energy density and pressure, respectively, and $g^{\mu\nu}$ is the metric tensor.
with $z$ oriented along the beam axis.
Below we use the curvilinear light-cone
coordinates $(\tau,x,y,\eta)$, where $\tau = \sqrt{t^2 - z^2}$ is the proper time and \mbox{$\eta=\frac{1}{2} \ln \frac{t+z}{t-z}$}~ is the space-time rapidity.
In the case of the longitudinal boost--invariant (2+1)--dimensional flow one can represent
the fluid's four-velocity as~\cite{Kaj86}
$u^{\mu}=\gamma_{\perp} (\cosh{\eta}, {\bm v}_{\perp}, \sinh{\eta})$,
where \mbox{${\bm v}_{\perp}$} is the transverse velocity in the symmetry
plane \mbox{$z=0$} and \mbox{$\gamma_{\perp} = (1 - v_{\perp}^2)^{-1/2}$} stands for the
transverse Lorentz factor.
To solve Eq.~(\ref{hydro}) one needs the equation of state (EoS), namely
a relation connecting $P$ and~$\varepsilon$. For chemically nonequilibrated matter considered in this paper,  $P=P\hsp (\varepsilon,\lambda)$, where $\lambda$ is the (anti)quark fugacity. In this work we use a simple analytic parametrization for~$\lambda$ as a~function of the proper time 
(see below~\re{lambda}).

Additionally, we calculate the space-time dependence of the local proper time $\tau_P$ of a fluid cell element by solving the equation
\bel{taup}
u^{\mu} \partial_{\mu} \tau_P ~ = ~ 1~,~~~~~~
\tau_P(\tau_0,x,y,\eta) ~=~ \tau_0 ~,
\ee
where the parameter $\tau_0$ corresponds to initial longitudinal proper time of the hydrodynamic expansion.
Equation~(\ref{taup}) is solved simultaneously with \re{hydro}.
In general, $\tau_P$ is found to be smaller than the 'global' time $\tau$ due to the presence of non-zero transverse flow.
In the limiting case of the one-dimensional longitudinal Bjorken expansion \cite{Bjo83}, one has $v_{\perp}=0$ and, consequently, $\tau_P=\tau$.

\begin{figure}[ht]
\centering
\includegraphics[width=0.49\textwidth]{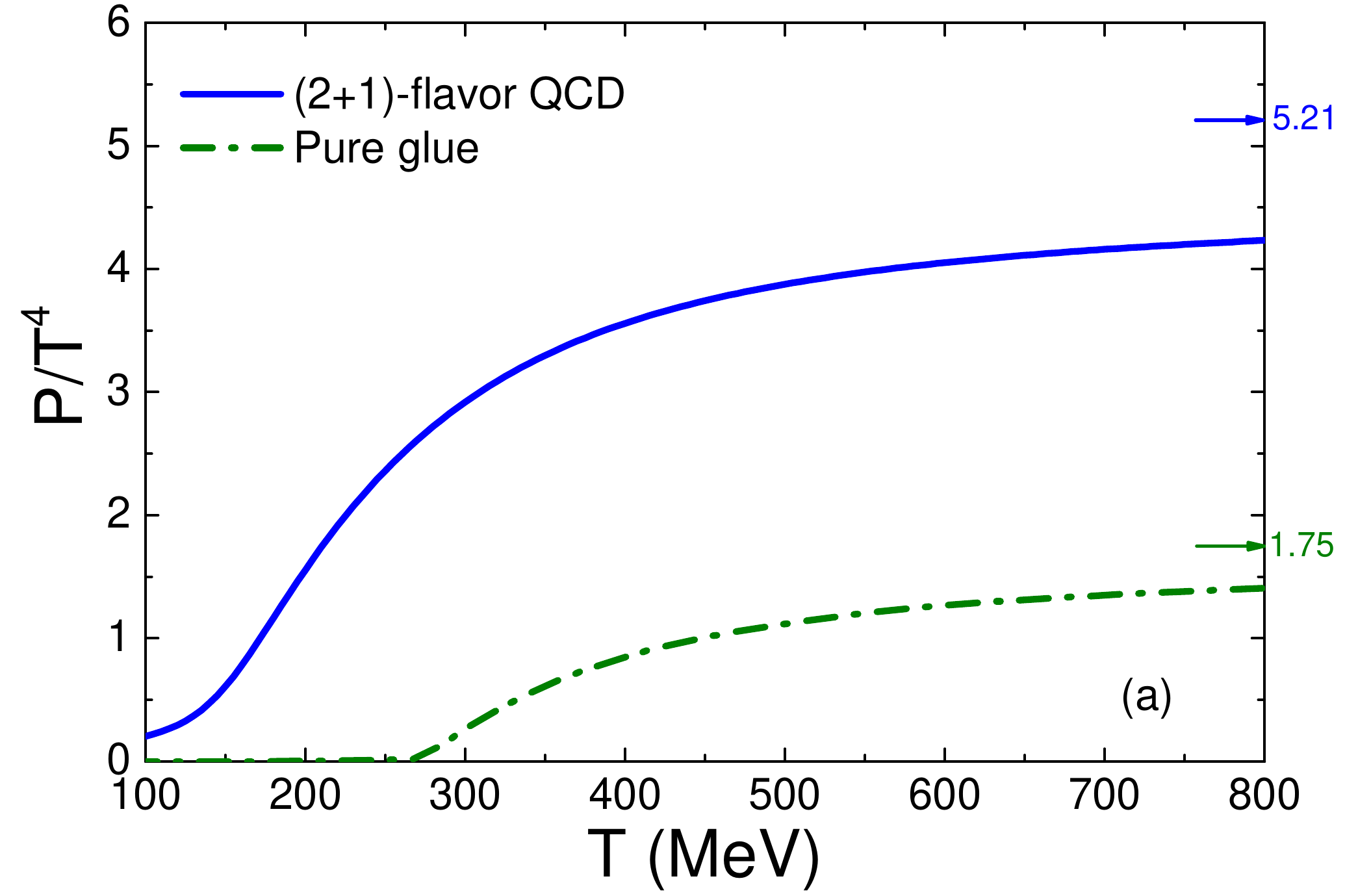}
\includegraphics[width=0.49\textwidth]{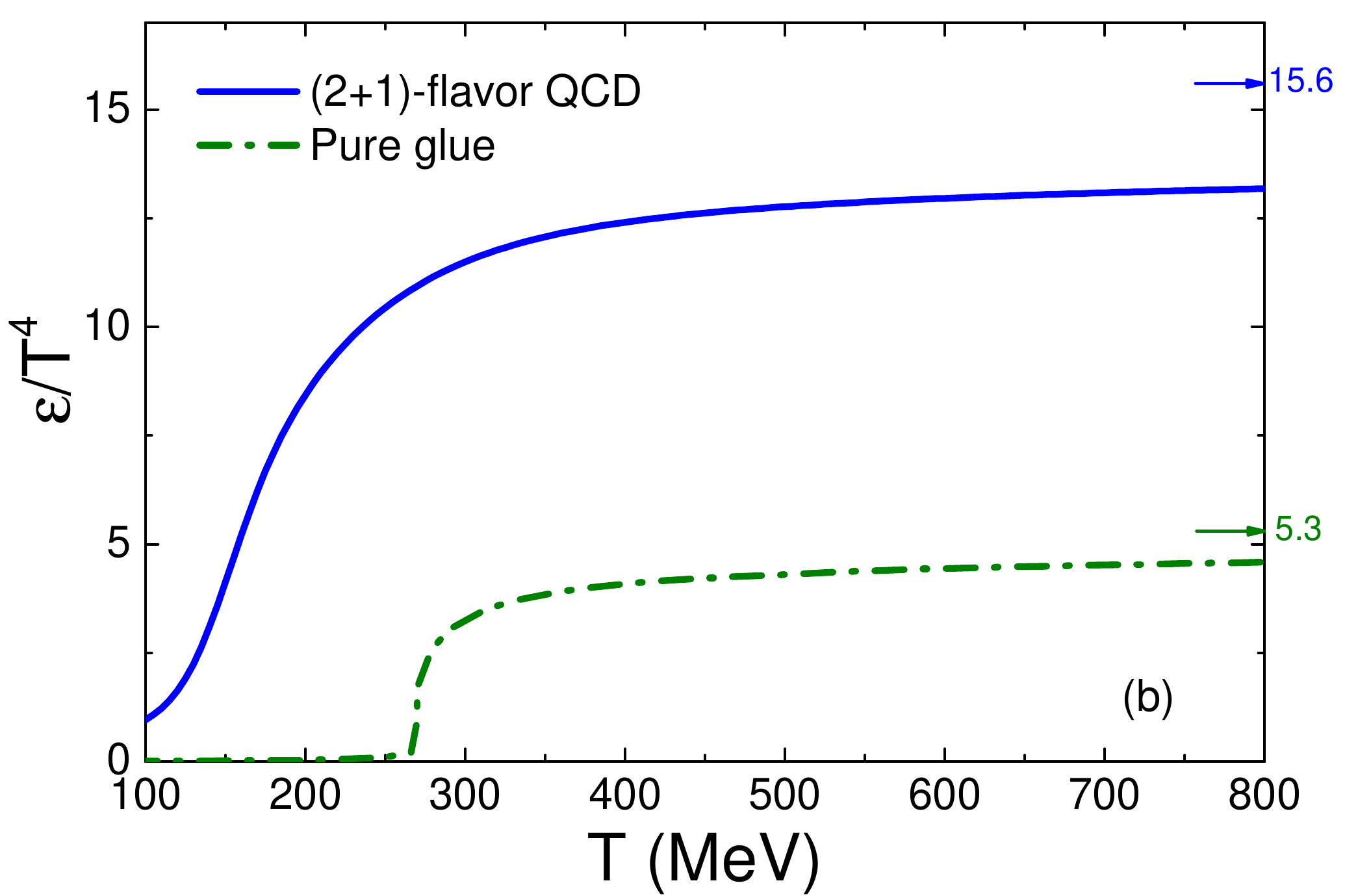}
\caption[]{
Temperature dependence of the scaled pressure~(a) and the scaled energy density~(b)
from the lattice QCD calculations~\cite{WBQCD,WBSU3}.
}\label{fig:latticeEoS}
\end{figure}

In order to construct the equation of state we
use the lattice {QCD} calculations for the EoS of the {strongly interacting}
matter in two limiting cases: 1)~{the} chemically equilibrated {QCD matter}~\mbox{\cite{WBQCD,HQCD}},
2) the SU(3) gluodynamics without (anti)quarks~\cite{Boy95,WBSU3}. {In the following we denote}
these cases as FQ (Full QCD) and PG (Pure Glue), respectively.
The FQ case corresponds to the (2+1)-flavour QCD calculations
which predict the crossover-type
transition at $T\sim 155~\textrm{MeV}$.
The PG calculation provides a first-order deconfinement phase transition at $T=T_c\simeq 270~\textrm{MeV}$.
The temperature dependencies of the pressure and energy density for FQ and PG scenarios
are depicted in Fig.~\ref{fig:latticeEoS}.
Larger values of $P$ and~$\varepsilon$ in the FQ calculation appear due to the contribution
of quark-antiquark degrees of freedom. Note that there is a discontinuity of $\varepsilon(T)$
at $T=T_c$ in the PG case. We note that very small values of $P$ and $\varepsilon$ at $T<T_c$ in the PG matter originate from large masses of glueballs ($M_g\gg T_c$)
which are
the constituents of the confined phase~\cite{WBSU3}.

The suppression of the quark and antiquark densities
at given temperature
is characterized by the (anti)quark fugacity $\lambda$ (for details, see Ref.~\cite{Vov16}).
Generalizing the lattice EoS for the chemically nonequilibrium case
with $\lambda<1$ is not a~straightforward task. We construct the $P$ and $\varepsilon$ as functions of $T$ and $\lambda$ by using the
linear interpolation (LI) between the the PG and FQ cases:
\begin{eqnarray}
P\,(T,\lambda) &=& \lambda \, P_{\hsp\rm FQ}\hsp (T) + (1 - \lambda) \, P_{\hsp\rm PG}\hsp (T)\hsp ,
 \label{eq:pint} \\
\varepsilon\,(T,\lambda) &=& \lambda \, \varepsilon_{\hsp\rm FQ}\hsp (T) + (1 - \lambda) \,
\varepsilon_{\hsp\rm PG}\hsp (T)\hsp .
\label{eq:eint}
\end{eqnarray}

After excluding the temperature variable in Eqs.~(\ref{eq:pint})--(\ref{eq:eint}), one gets the
relation $P=P(\varepsilon,\lambda)$ which is  used in  hydrodynamic simulations.
The limits $\lambda=0$ and $\lambda=1$ correspond to the thermodynamic functions of the PG
and FQ matter, respectively. Note that the linear \mbox{$\lambda$--\hsp dependence} of $P$ and $\varepsilon$ is a characteristic
feature of the ideal gas of massless gluons and (anti)quarks studied in Refs.~\cite{Sto15,Vov16}.

Using Eqs.~(\ref{eq:pint}) and (\ref{eq:eint}) and basic thermodynamic identities,
one can calculate the total density of (anti)quarks $n_q$ and the entropy density $s$\hsp .
The following relations are obtained
\begin{eqnarray}
n_{\hsp q}\,(T,\lambda) &=& \frac{\lambda}{T} \, ( P_{\hsp\rm FQ}-P_{\hsp\rm PG} ),
\label{eq:nqint} \\
s\,(T,\lambda) &=& \lambda \, s_{\hsp\rm FQ} (T) + (1 - \lambda) \, s_{\hsp\rm PG} (T)
- n_{\hspm q} (T,\lambda)\hsp\ln{\lambda}\hsp .
\label{eq:sint}
\end{eqnarray}

We assume that at $\tau=\tau_0$ the initial (anti)quark
densities vanish in all cells and gluons are in thermal and chemical equilibrium.
Similarly to Refs.~\cite{Sto15,Vov16} we postulate that~$\lambda$ is an
explicit function of the local proper time $\tau_P$ which increases from $\lambda=0$ at $\tau_P=\tau_0$ to $\lambda=1$ at $\tau_P-\tau_0\rightarrow \infty$. The following simple parametrization is used:
\bel{lambda}
\lambda(\tau_P) = 1 - \exp\left( \frac{\tau_0-\tau_P}{\tau_*} \right),
\ee
where $\tau_*$ is a model parameter characterizing the chemical equilibration time of (anti)quarks.
{There are different} estimates for $\tau_*$ {in the literature} ranging from
\mbox{$\tau_*\sim 1~\textrm{fm}/c$~\cite{Rug15}} to \mbox{$\tau_*\sim 5~\textrm{fm}/c$~\cite{Xu05}}.
Note that $\tau_*=0$ corresponds to the instantaneous chemical equilibration of quarks and gluons.

\section{Hydrodynamic simulations for Yang-Mills and equilibrium QCD equations of state}

\subsection{CLVisc hydro simulation}

\begin{figure}[htb!]
\centering
\includegraphics[width=0.90\textwidth]{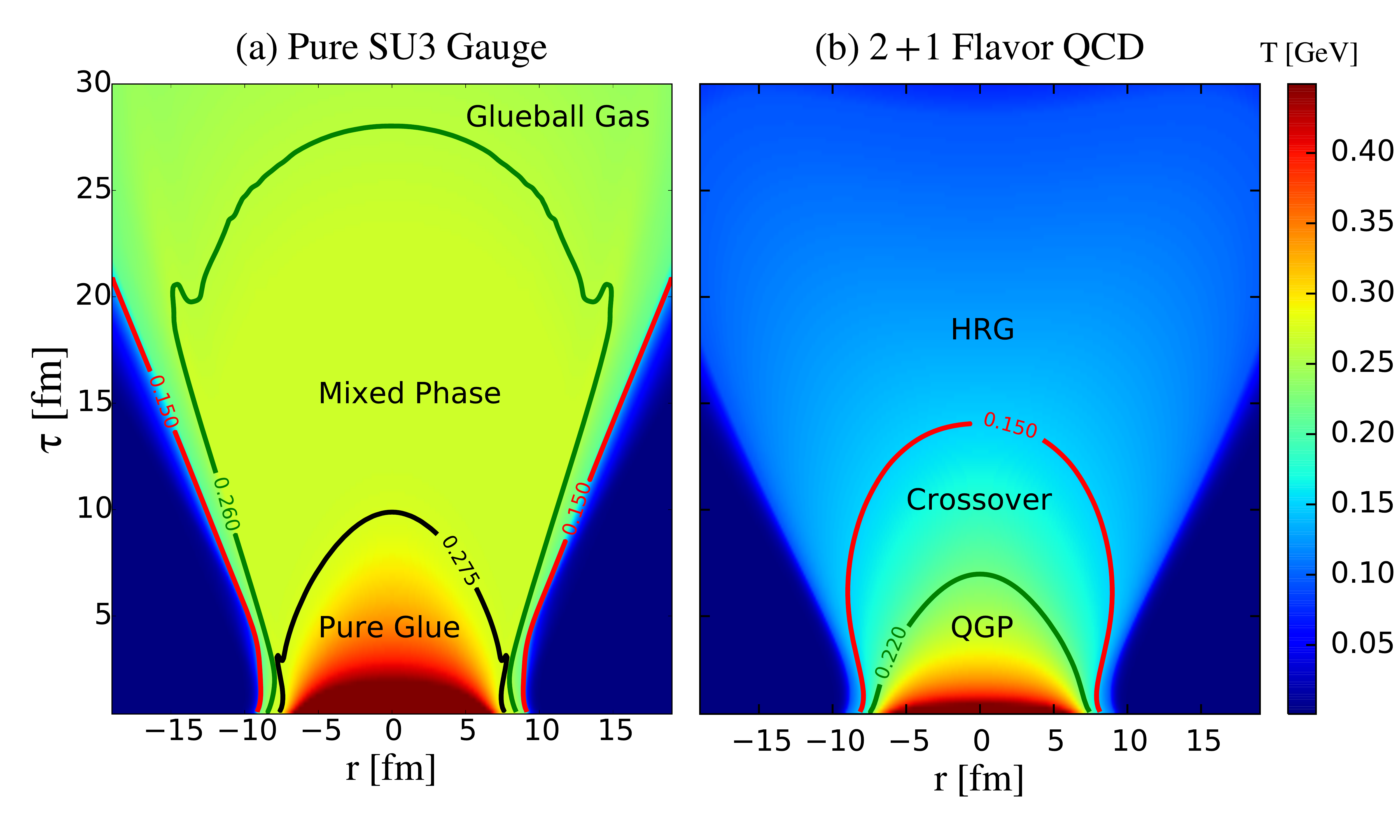}
\caption[]{
The temperature in the $r-\tau$ plane for an ideal hydrodynamic
simulation of the purely central Pb+Pb collisions within the CLVisc~\cite{clvisc} hydrodynamic code.
Here $r$ is the transverse coordinate in the reaction plane.
The initial entropy deposition at $\tau_0 = 0.4$~fm/$c$ in the transverse plane is proportional to the
number of wounded nucleons from an optical Glauber model.
The (a) Yang-Mills  and (b) (2+1)-flavor QCD  equations of
state obtained from lattice simulations are used.
}\label{fig:HydroLG}
\end{figure}

As a first step we perform the hydrodynamic simulations of heavy-ion collisions for two limiting cases: the (2+1)-flavor full QCD matter and the quarkless pure Yang-Mills (YM) matter. The former case corresponds to $\tau_* = 0$~fm/$c$ while the latter implies $\tau_* \to \infty$. Such a comparison is rather instructive and may indicate the possible effects of the pure glue scenario that one can later study in a more realistic approach.
For this purpose we use three different hydrodynamic codes.
First, we use
the CLVisc hydro code~\cite{clvisc} and simulate the central Pb+Pb collisions. The initial entropy deposition at $\tau_0 = 0.4$~fm/$c$ in the transverse plane is proportional to the
number of wounded nucleons calculated in the optical Glauber model.
The normalization constant is fixed in order to reproduce the maximum initial energy density of 166~GeV/fm$^3$, which is reached in the central cell.
The resulting space-time profile of the temperature in the $r-\tau$ plane is shown in Fig.~\ref{fig:HydroLG}. Here $r$ is the transverse coordinate in the reaction plane.

For the equation of state of the full QCD the resulting temperature profile, shown in the right panel of Fig.~\ref{fig:HydroLG} is rather typical for hydrodynamic simulation in ultrarelativistic $A+A$ collisions: the system smoothly and rather rapidly cools down to the ``freeze-out'' temperatures of $100-150$~MeV within the time interval of $10-15$~fm/$c$.
The situation, however, is very different for the pure YM scenario.
The system cools down to the critical temperature of $T_c = 270$~MeV initially, then spends a significant amount of time in the mixed phase of the deconfinement phase transition, and finally enters the phase of quickly decaying glueballs at $T \sim 250$~MeV. The resulting lifetime of the system appears to be much longer as compared to the full QCD case: it can be about 2-3 times larger and is a consequence of undergoing the long deconfinement phase transition during the system evolution.

\begin{figure}[htb!]
\centering
\begin{minipage}{.49\textwidth}
\centering
(a) Pure SU3 Gauge
\\
\includegraphics[width=\textwidth]{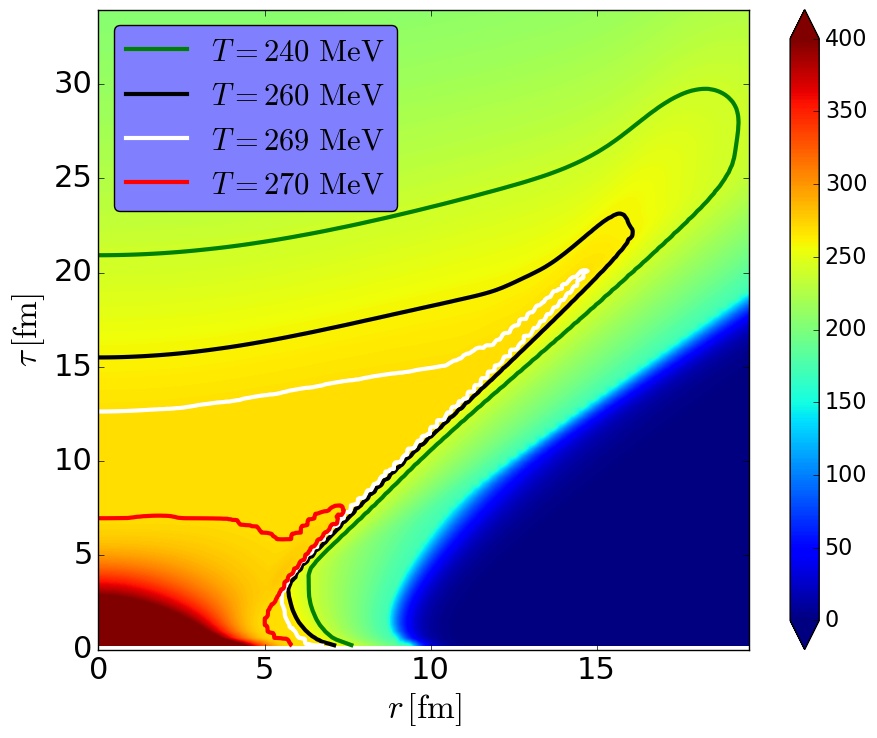}
\end{minipage}
\begin{minipage}{.49\textwidth}
\centering
(b) 2+1 Flavor QCD
\\
\includegraphics[width=\textwidth]{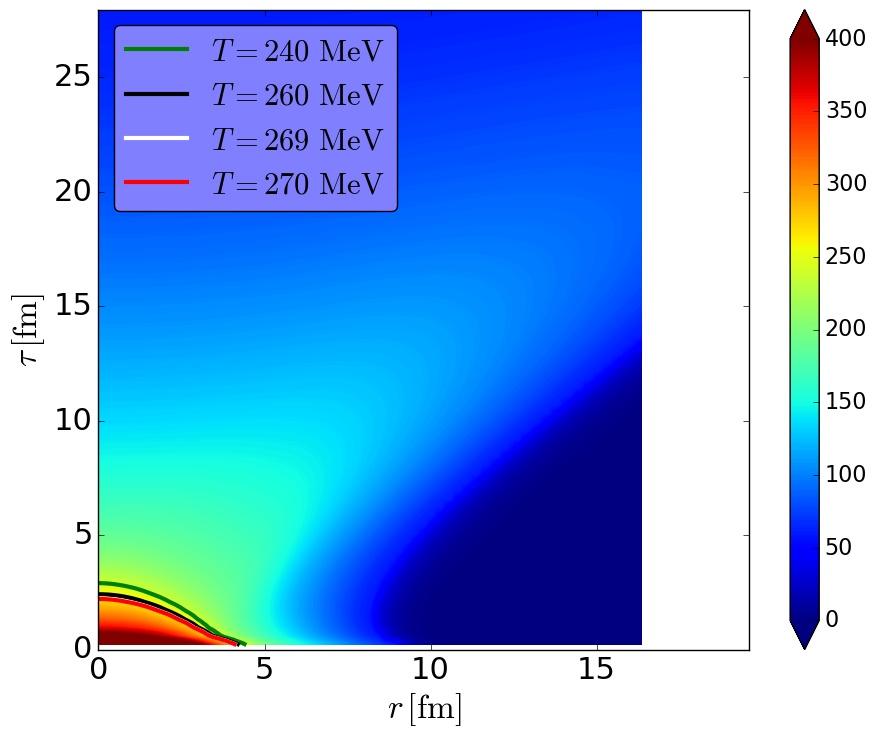}
\end{minipage}
\caption[]{
The temperature in the $r-\tau$ plane for an ideal hydrodynamic
simulation of the non-central Pb+Pb collisions within the SHASTA-based hydrodynamic solver introduced in Refs.~\cite{Molnar:2009tx, Niemi:2012ry}. 
In both cases the initial energy density profile 
is proportional to the density of binary nucleon collisions, but the normalization is fixed such that the final pion multiplicity is approximately the same in both calculations. The impact parameter $b = 7.74$ fm corresponds to the $20-30$ \% centrality class.
The (a) Yang-Mills and (b) (2+1)-flavor QCD  equations of
state based on lattice simulations are used.
}\label{fig:Harri-T}
\end{figure}

\begin{figure}[htb!]
\centering
\begin{minipage}{.49\textwidth}
\centering
(a) Pure SU3 Gauge
\\
\includegraphics[width=\textwidth]{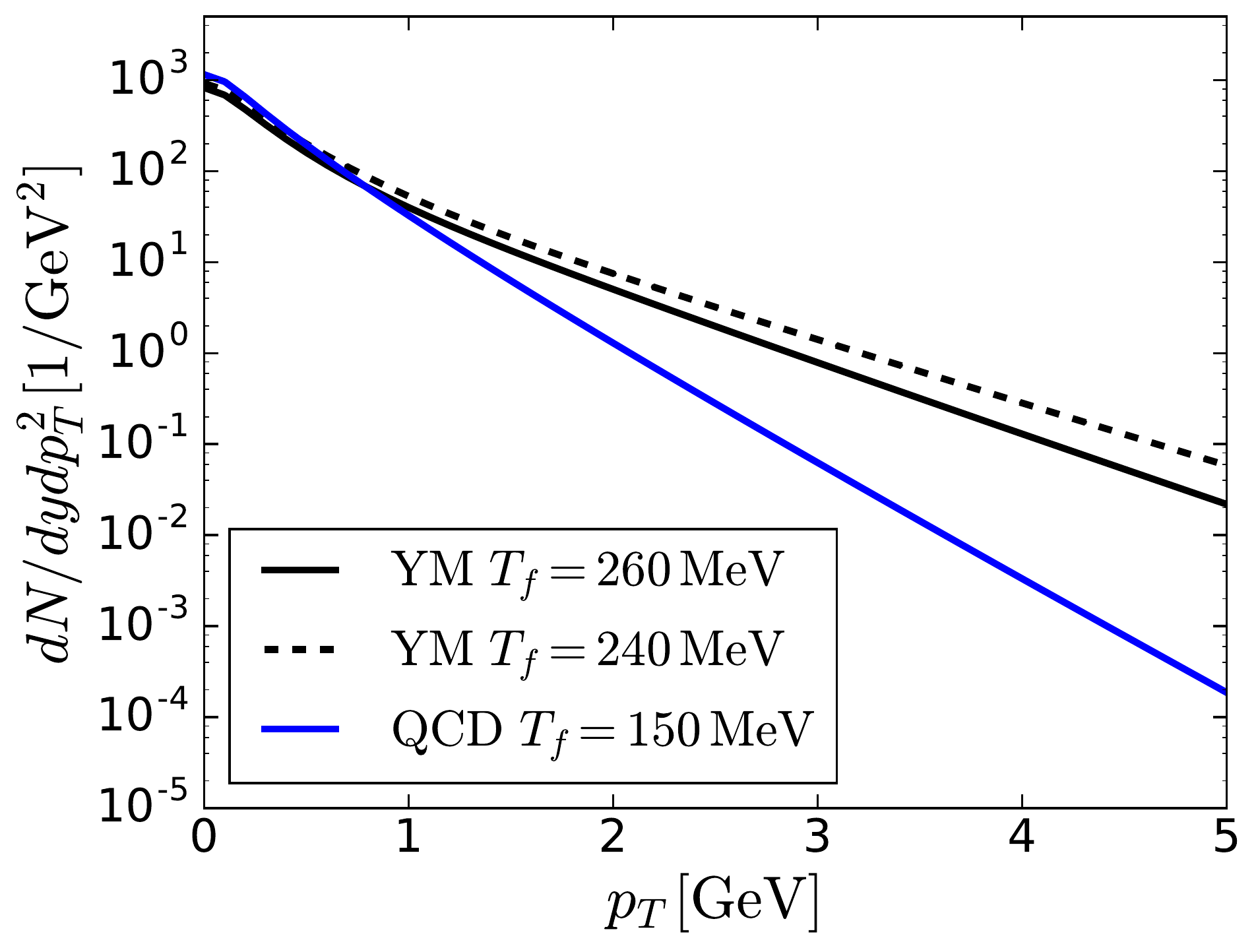}
\end{minipage}
\begin{minipage}{.49\textwidth}
\centering
(b) 2+1 Flavor QCD
\\
\includegraphics[width=.96\textwidth]{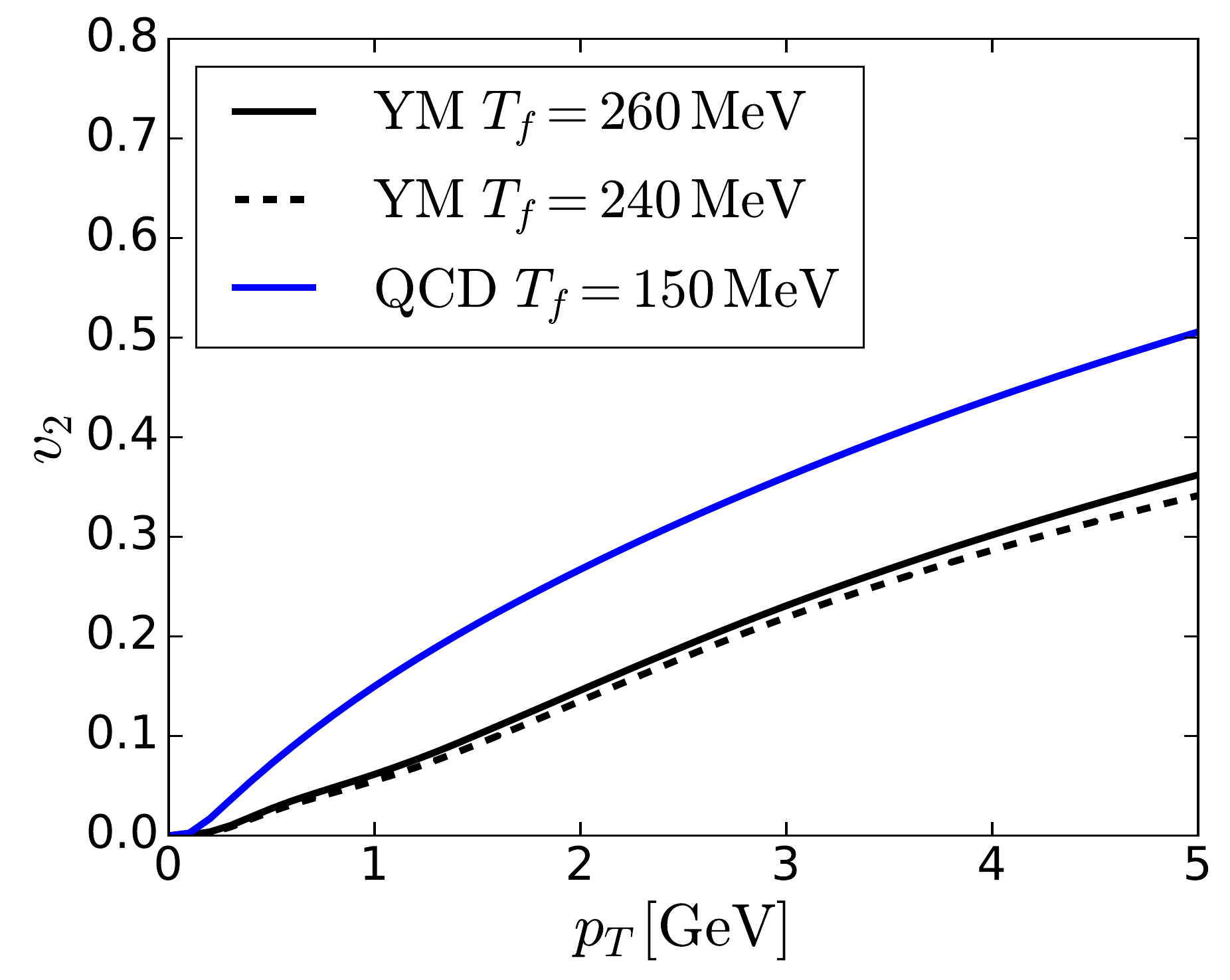}
\end{minipage}
\caption[]{
Spectra (a) and elliptic flow (b) of the positively charged pions using the pure Yang-Mills or full QCD equation of state. The decoupling
temperatures, that had been taken to be close to the corresponding QCD/Yang-Mills transition temperature, are indicated in the legends.
}\label{fig:Harri}
\end{figure}

\subsection{SHASTA-based hydro simulation}

A similar calculation of the temperature profile performed within the SHASTA-based
hydrodynamical solver introduced in Refs.~\cite{Molnar:2009tx, Niemi:2012ry} is shown in Fig.~\ref{fig:Harri-T},
where the hydrodynamical calculation using the pure YM EoS is compared to the one using a parametrization~\cite{Huovinen:2009yb} of (2+1) flavor QCD EoS. 
For this calculation the initial energy density profile 
is taken to be proportional to the density of binary nucleon collisions, but the normalization is fixed such that the final pion multiplicity is approximately the same in both the YM and full QCD calculations. The impact parameter is $b = 7.74$ fm, which roughly corresponds to the $20-30$\% centrality class. Despite some differences in the initial conditions the temperature profiles shown in Fig.~\ref{fig:Harri-T} are generally consistent with the ones depicted in Fig.~\ref{fig:HydroLG}, which were obtained with the CLVisc hydro code.
In Fig.~\ref{fig:Harri}a we show the transverse momentum spectra and in Fig.~\ref{fig:Harri}b the elliptic flow coefficient $v_2$ of the positively charged pions for both pure Yang-Mills and the (2+1) flavor QCD equations of state.
The pion spectra in the pure SU(3) 
case are calculated by constructing an equation of state of free (non-interacting) glueballs, such that pressure as a function of temperature, as well as its slope 
is approximately the same around $T \sim 250$ MeV in both pure Yang-Mills and glueball EoS. The number of glueball states is the same as listed 
in Ref.~\cite{Meyer:2004gx}, but the masses of the states are quite freely modified in order to get a good match to the pure YM EoS. 
The glueball states then decay into pions and kaons.

\subsection{vHLLE hydro simulation}

\begin{figure}[h!]
\centering
\begin{minipage}{.49\textwidth}
\centering
(a) Pure SU3 Gauge
\\
\includegraphics[width=\textwidth]{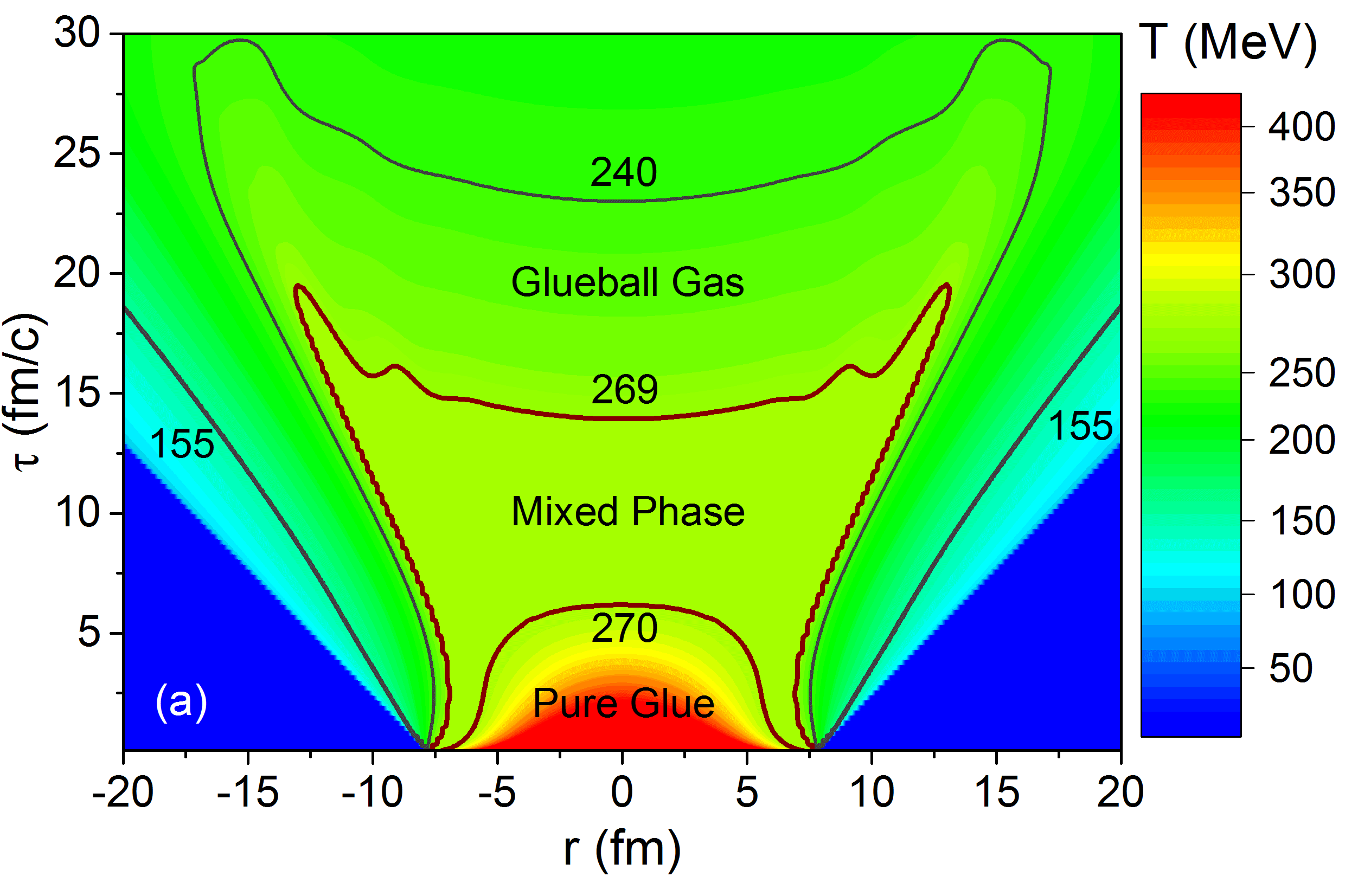}
\end{minipage}
\begin{minipage}{.49\textwidth}
\centering
(b) 2+1 Flavor QCD
\\
\includegraphics[width=\textwidth]{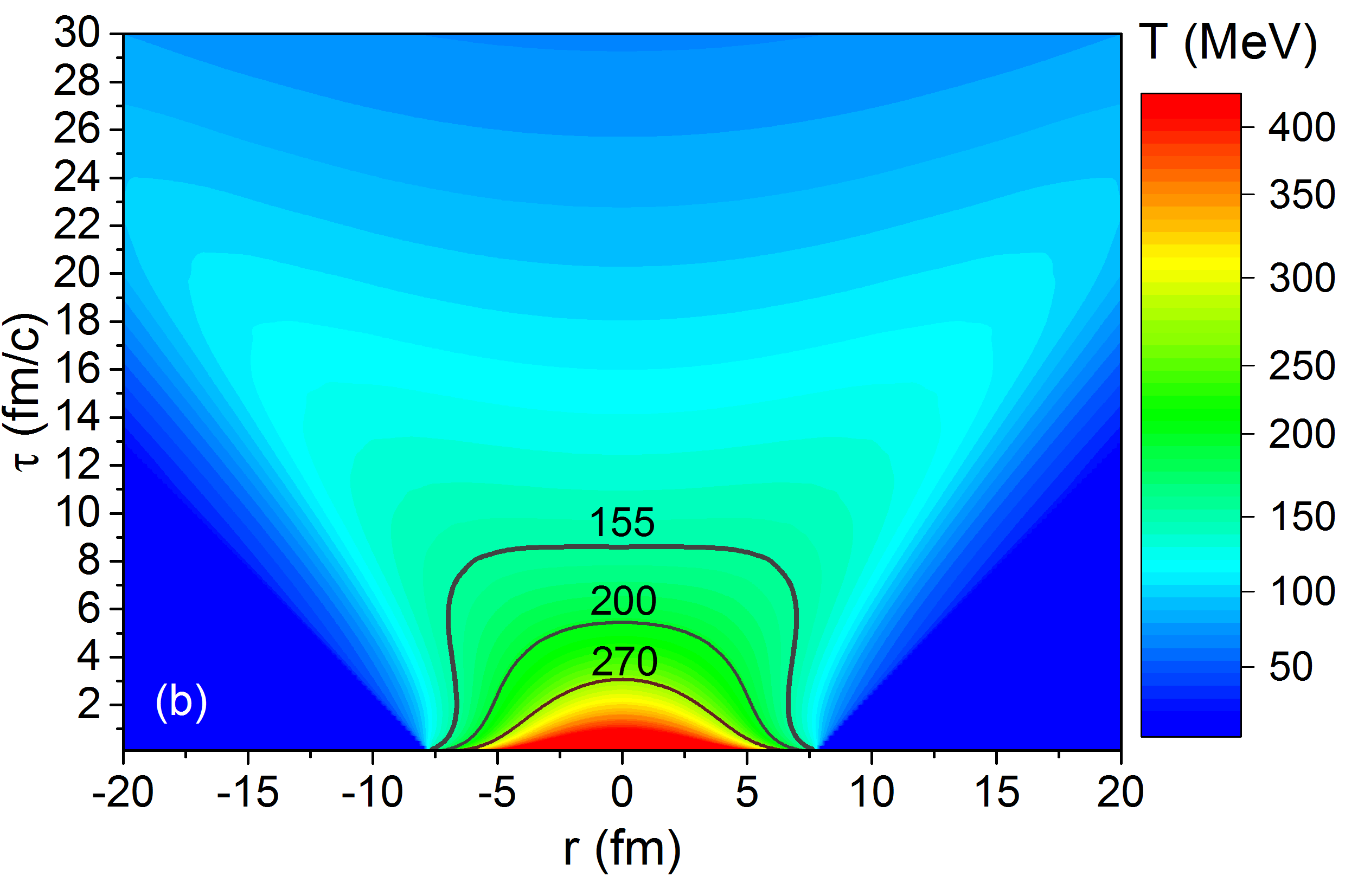}
\end{minipage}
\caption[]{
The temperature in the $r-\tau$ plane for an ideal hydrodynamic
simulation of the 0-40\% central Pb+Pb collisions within the vHLLE hydrodynamic code~\cite{vHLLE}. In both cases the initial energy density profile 
is proportional to the linear combination of the transverse distributions of wounded nucleons and of binary collisions taken from the
event-averaged Monte Carlo Glauber model with the same normalization constant.
The (a) Yang-Mills  and (b) (2+1)-flavor QCD  equations of
state obtained from lattice simulations are used.
}\label{fig:vhlle}
\end{figure}

It is also instructive to perform a comparable hydrodynamic simulation of  heavy-ion collisions with the YM and the (2+1)-flavor QCD EoS within another hydro code, namely the vHLLE hydrodynamic solver~\cite{vHLLE}.
The resulting temperature profiles in the 0-40\% central Pb+Pb collisions are shown in Fig.~\ref{fig:vhlle}. One can compare these calculations with the corresponding results from the two other codes shown in Figs.~\ref{fig:HydroLG} and \ref{fig:Harri-T}. While the initial conditions are slightly different in these three simulations they all show a consistent physical picture: the evolution of the YM matter in heavy-ion collisions is very different from that of a fully equilibrated QCD matter. The YM matter evolves for a much longer time and spends a significant portion of the space-time evolution in the region of mixed phase.

\section{Numerical results for PbPb collisions at LHC with pure glue initial state}

In a more realistic scenario one has to take into account the gradual appearance of (anti)quarks during the system evolution, and also retain the consistence with the available experimental data.
We model the Pb+Pb collisions at the LHC
with the {c.\hsp m.} energy per nucleon pair $\sqrt{s_{NN}} = 2.76$~TeV.
In our calculations we {choose} $\tau_0 = 0.1$~fm/$c$ as the initial longitudinal proper time of {the hydrodynamic} evolution.
It is assumed that there is no initial
transverse flow, i.e., \mbox{$\bm{v}_{\perp}(\tau_0,x,y)=0$}, and that the initial energy density profile
is proportional to the linear combination of the transverse distributions of wounded nucleons and of binary collisions taken from the
event-averaged Monte Carlo Glauber model as implemented in the GLISSANDO code~\cite{GLISSANDO}.
The coefficient of proportionality in the initial $\varepsilon$-profile
is fixed to reproduce the {observed} hadron spectra {within} the simulation assuming \textit{chemical
equilibrium} with the full QCD EoS for a~corresponding centrality interval~(see Ref.~\cite{KarpenkoLHC} for details). We use the same initial energy density profile in the present calculations for the chemical nonequilibrium case.
It is also assumed that the initial state is purely gluonic, i.e. that initially the fugacity $\lambda$ of (anti)quarks is zero, and that gluons are in chemical equilibrium at $\tau \geq \tau_0$.

Equations (\ref{hydro}) and (\ref{taup}) are solved using the (2+1)--dimensional
version of the vHLLE hydrodynamic code~\cite{vHLLE}.
We consider the 0\hspm --20\hsp\%
and 20\hspm --40\hsp\% central Pb+Pb collisions at LHC.
\begin{figure}[htb!]
\centering
\includegraphics[width=0.49\textwidth]{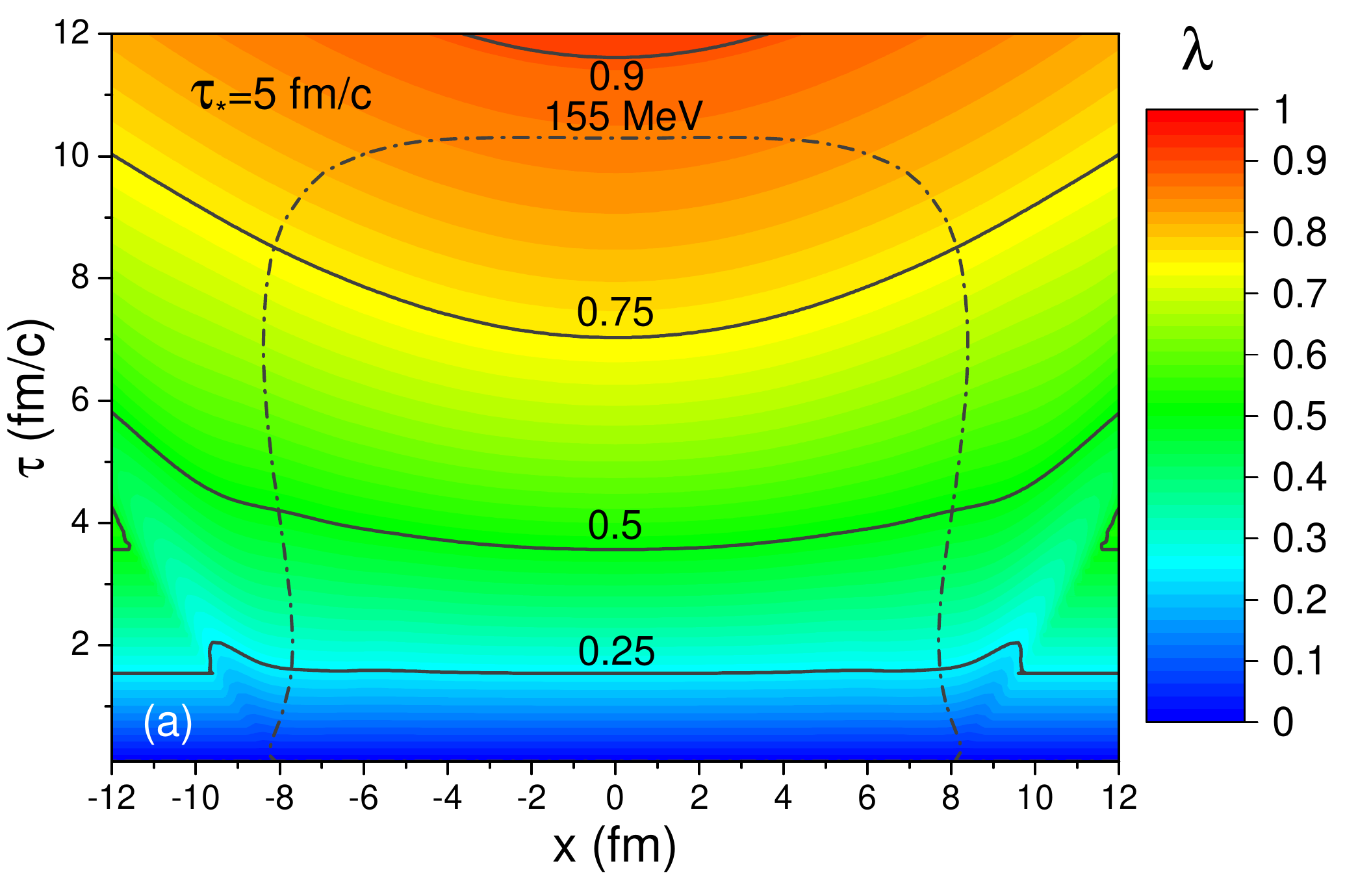}
\includegraphics[width=0.49\textwidth]{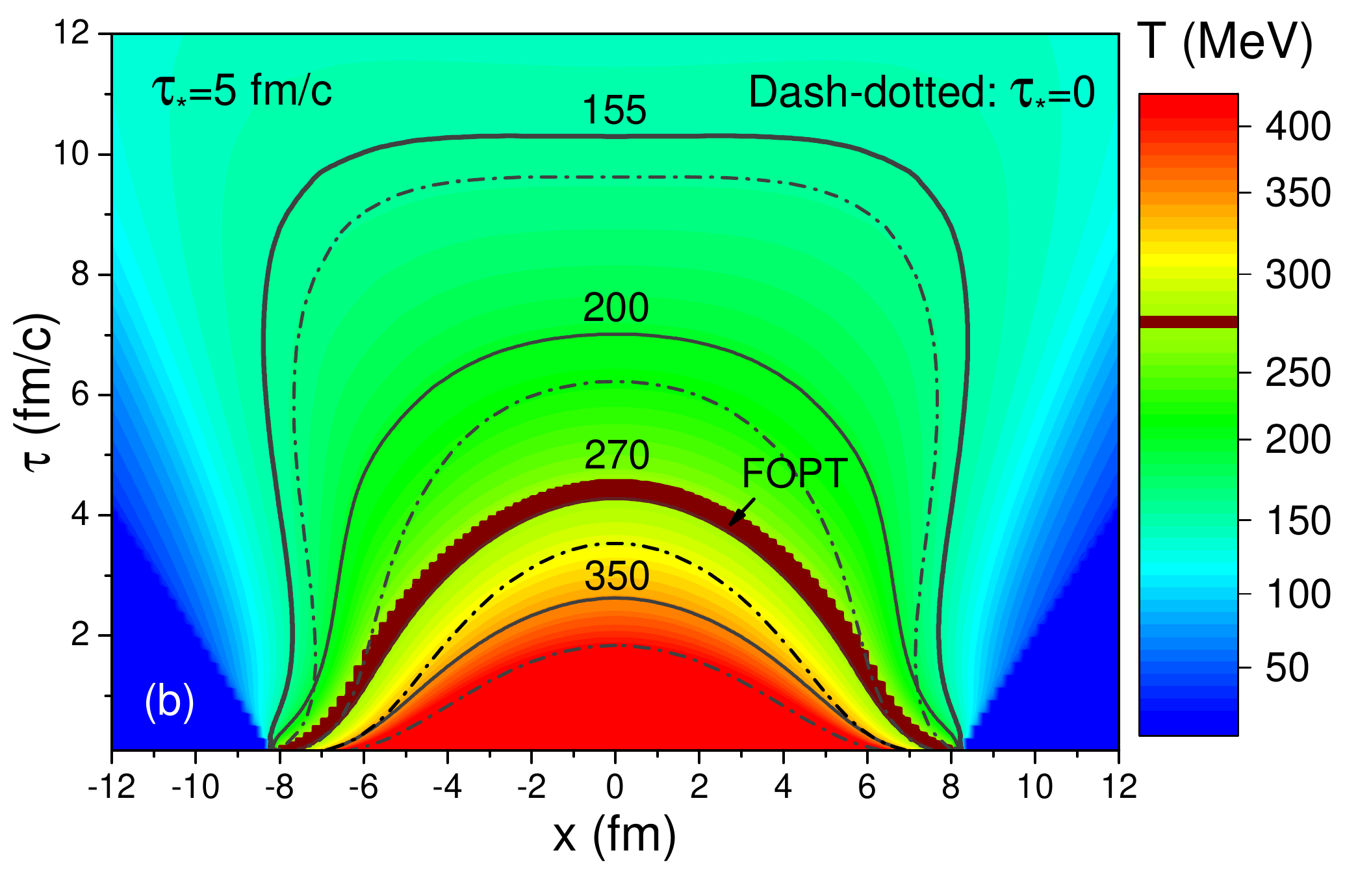}
\caption[]{
The density plots of the {quark fugacity (a) and temperature (b)} in the $x-\tau$ plane for the \mbox{0--20\hspm\%}
most central Pb+Pb collisions at \mbox{$\sqrt{s_{NN}} = 2.76$~TeV}.
The solid curves show {contours of~$\lambda$ and $T$}.
The~dark red region labeled by FOPT corresponds to the mixed-phase region of the first-order phase
transition at \mbox{$T=T_c\simeq~270~\textrm{MeV}$}. The dash-dotted curves in (b) depict isotherms calculated for
equilibrium QCD matter with $\lambda=1$.
}\label{fig:T}
\end{figure}

The contour plot of the quark fugacity $\lambda$ in the $x-\tau$ plane
is given in~Fig.~\ref{fig:T}\hsp a. The dashed line
shows the isotherm $T=155~\textrm{MeV}$ which presumably corresponds to the hadronization
hypersurface.
One can see that typical lifetimes of the deconfined
phase in the considered reaction do not exceed $10~\textrm{fm}/c$\hsp.
In Fig.~\ref{fig:T}\hsp a one observes that deviations from chemical equilibrium
($\lambda\lesssim 0.9$) may survive up to the hadronization {stage}.
As discussed in Ref.~\cite{Vov16} this may lead to a suppression
of (anti)baryon-to-pion ratios observed~\cite{Abe13} for the considered reaction.
Note that $\lambda$ evolves with $\tau$ in Fig.~\ref{fig:T}\hsp a even in the large $x$ regions where there is practically no matter. This behavior follows from applying \re{lambda}. In reality, of course, the values of $\lambda$ for these very dilute and cold  fluid elements
are irrelevant and should be ignored.

Figure ~\ref{fig:T}\hsp b shows the contour plot of the temperature in the plane $(x,\tau)$.
The solid and dashed curves correspond to
$\tau_*=5~\textrm{fm}/c$ and $\tau_* = 0$, respectively.
One {can see} that the chemically undersaturated matter is hotter as compared to
the equilibrium case (\mbox{$\lambda=1$}),
especially at the earlier times\hsp\footnote
{
Note that in both cases we take the same profile of the energy density at $\tau=\tau_0$.
}. This is a consequence of reduced
number of degrees of freedom in such a medium. According to Fig.~\ref{fig:T}\hsp b,
typical lifetimes of the mixed phase are rather short, they do not
exceed $0.5~\textrm{fm}/c$\hsp . This is at variance with calculations in the (1+1) dimensional hydrodynamics which predict~\cite{Sat07} much larger lifetimes of the mixed phase within the chemically equilibrated bag model.
Our nunerical results show that about 25\% of the total final entropy is generated during the ideal hydro evolution of the chemically undersaturated QCD matter, in line with results of the semi-analytic calculation in Ref.~\cite{Vov16}.

\section{Direct photon and thermal dilepton emission at LHC\label{dpes}}

The emission of direct\hspm\footnote
{
By direct photons we denote the 'non-cocktail' photons i.e.~those which are not produced in decays of
$\pi^0,\eta,\rho,\eta^\prime$, and $\phi$ mesons in the final stage of the reaction.
}
photons from expanding matter created in relativistic A+A collisions
has several components~\cite{Paq15,Lin16}: a)~'prompt' photons from binary collisions of initial nucleons, b)~'thermal' photons
from the high-temperature deconfined phase, c)~direct photons from {the} low-temperature hadronic phase.
The contribution of prompt photons becomes dominant at large transverse momenta.
As we will see below, this greatly reduces the~sensitivity
of photon $p_{\hsp T}$-spectra to chemical nonequilibrium effects.
However, the situ\-ation with transverse flows of photons
is different because of low azimuthal anisotropy of prompt photons.
Note that the \mbox{ALICE} experiments~\cite{Loh13} reveal
large elliptic flows of direct photons, which still can not
be explained in the chemically equilibrium scenario~\cite{Paq15}.

\begin{figure*}[hbt!]
\centering
\includegraphics[width=0.49\textwidth]{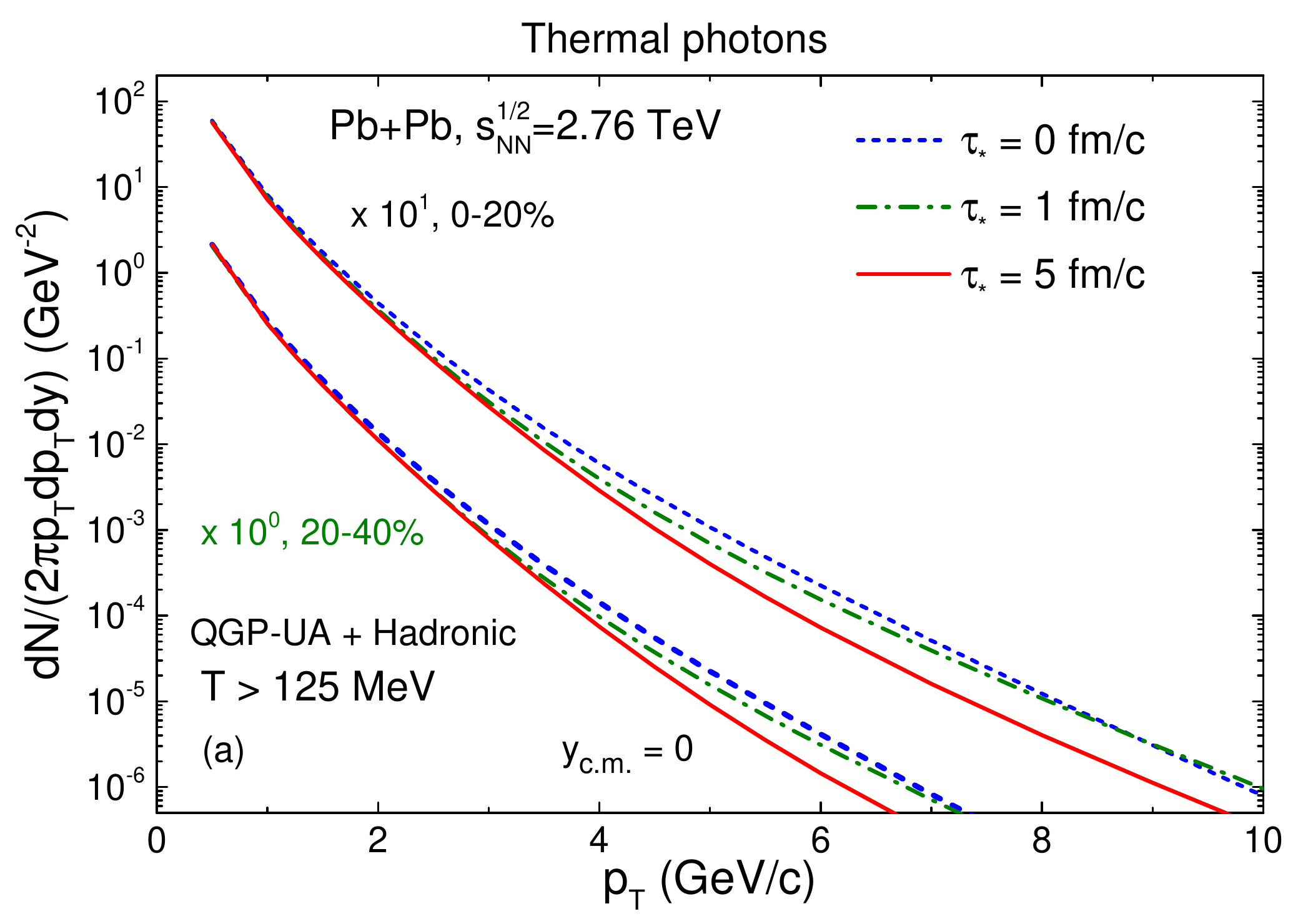}
\includegraphics[width=0.49\textwidth]{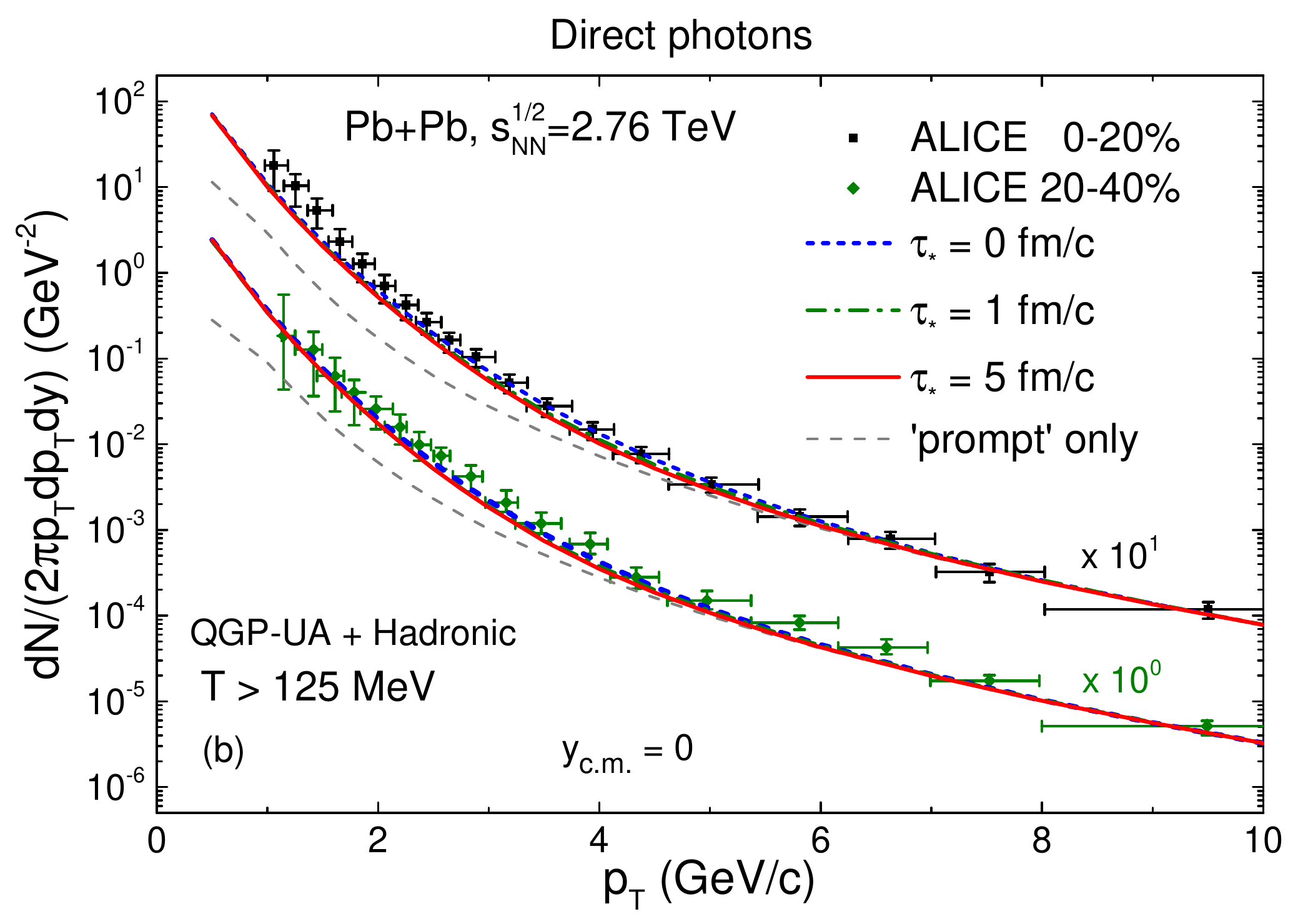}
\caption[]{
Spectra of thermal~(a) and direct~(b) photons in {the 0--20\hspm\%}
and 20-40\%
central Pb+Pb collisions at $\sqrt{s_{NN}}=2.76~\textrm{TeV}$ calculated
with the cutoff temperature of $125~\textrm{MeV}$.
The dashed, dash-dotted, and solid curves
correspond to $\tau_*=0\hspm , 1$, and $5~\textrm{fm}/c$\,, respectively. Dots with error bars show the experimental data~\cite{ALICEphot}.
}
\label{fig:phspec}
\end{figure*}
%--------------------------------------------------------------------

Figure~\ref{fig:phspec}\hsp{a}
shows our results for the thermal photon spectrum in {the 0--20\hspm\%}
and 20--40\hspm\%
central Pb+Pb collisions at $\sqrt{s_{NN}}=2.76~\textrm{TeV}$
calculated with the cut-off temperature of $T_f=125$~MeV.
We apply the parametrization QGP-UA~\cite{Vov16a} of the
photon production rate in a chemically nonequilibrium QGP. 
The low-$p_T$ spectrum looks very similar in all scenarios, while there is a sizable difference at high $p_T$.
The spectra of direct photons, i.e.
those that additionally include the prompt photons,
is depicted in Fig.~\ref{fig:phspec}\hsp{b}.
The inclusion of prompt photons makes the difference between different scenarios at high $p_T$ rather small.

The direct photon production in Pb+Pb collisions at LHC has been considered in various theoretical models
(\cite{Chatterjee12,Hees15,Paq15,Lin15}.
As noted in Ref.~\cite{ALICEphot}, the present uncertainties in the heavy-ion photon data at LHC do not allow to conclusively discriminate between the various
scenarios.

\begin{figure*}[hbt!]
\centering
\includegraphics[width=0.49\textwidth]{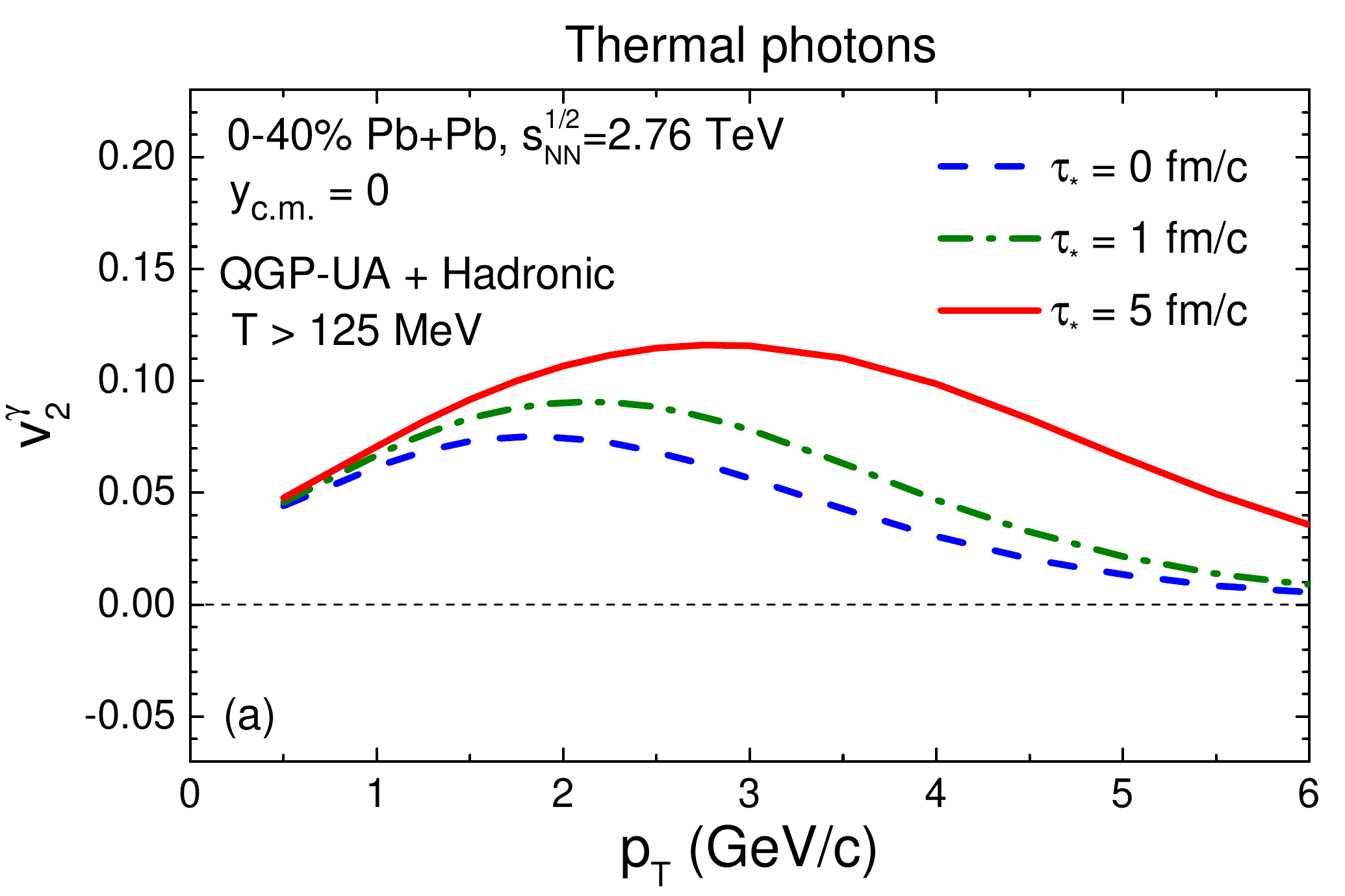}
\includegraphics[width=0.49\textwidth]{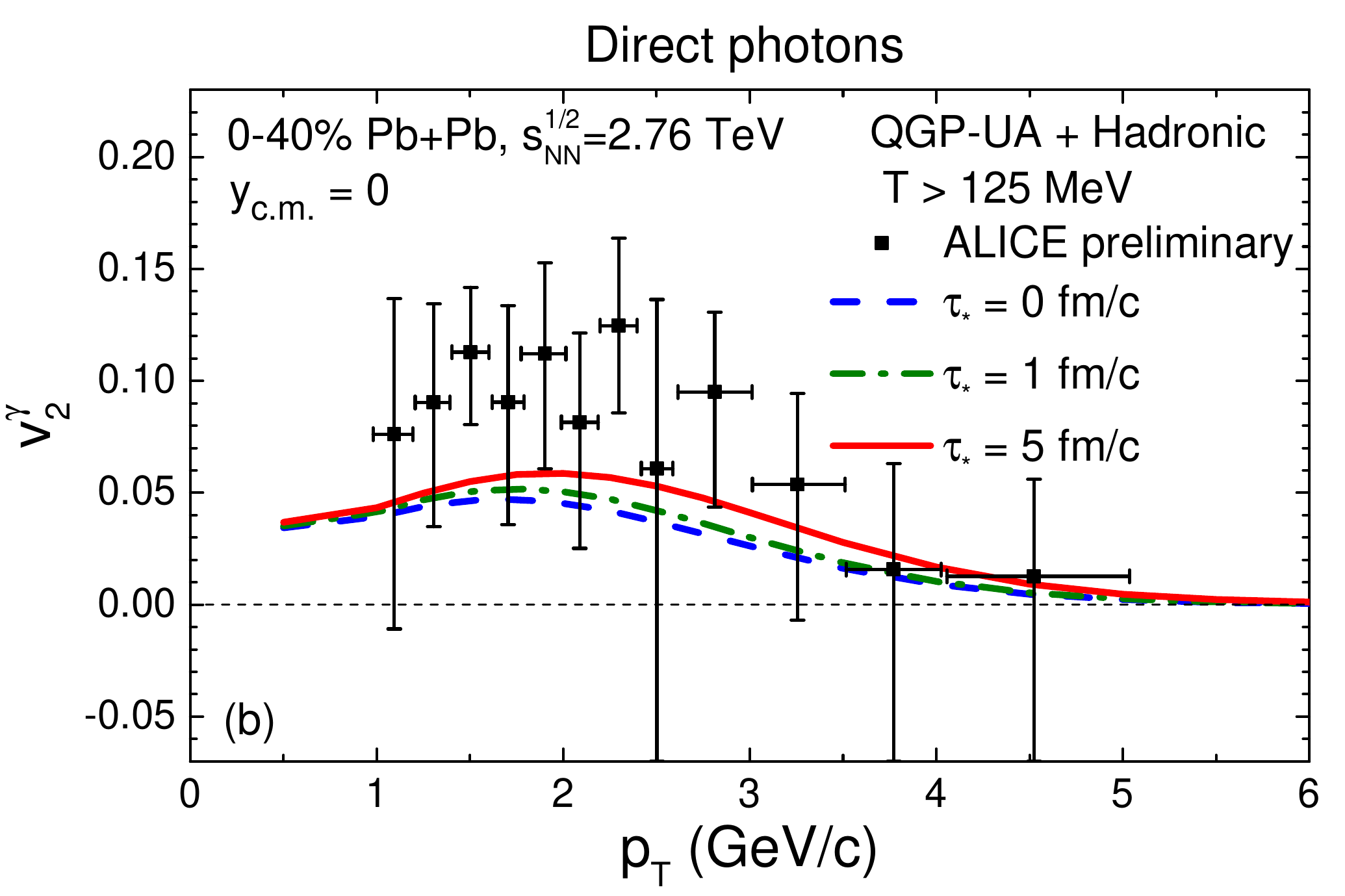}
\caption[]{
%(Color online)
Elliptic flow $v_{\hspm 2}^{\gamma}$ of thermal (a) and direct (b) photons as a function of transverse momentum $p_{\hsp T}$
in the~~$0-40$\hspm\%
central Pb+Pb collisions at $\sqrt{s_{NN}}=2.76~\textrm{TeV}$ calculated
with the cutoff temperature \mbox{$T_f=125$}~MeV.
The dashed, dash-dotted, and solid lines
correspond to $\tau_*=0\hsp , 1$ and $5~\textrm{fm}/c$\,, respectively\hspm .
Experimental data are taken from Ref.~\cite{Loh13}.
}
\label{fig:phv2}
\end{figure*}
%--------------------------------------------------------------------
The photon elliptic flow
$v_{\hspm 2}^{\gamma}(p_{\hsp T})$ is calculated by
\bel{elfl}
v_{\hspm 2}^{\gamma}(p_{\hsp T}) = \frac{\int_0^{2\hspm\pi} d\hsp\varphi \,
\frac{d N_{\gamma}}{d^{2} p_{\hsp T} dY} \, \cos(2\varphi)}{\int_0^{2 \pi} d\hsp\varphi\,\frac{d N_{\gamma}}{d^{2} p_{\hsp T} dY}}\,.
\ee

The photon spectrum, entering this equation includes both thermal and prompt components. We assume that prompt photons are azimuthally symmetric. Therefore, they contribute only to the denominator of~\re{elfl} reducing
$v_{\hspm 2}^{\gamma}$ at large $p_{\hsp T}$. The results of the calculations of the photon elliptic flow are shown in Fig.~\ref{fig:phv2}. 
In the pure-glue scenario the momentum anisotropy is significantly enhanced for the high-$p_T$ thermal photons. As seen from Fig.~\ref{fig:phv2}\hsp b, the inclusion of prompt photons notably decreases the effect.

\vspace*{2mm}
\begin{figure}[htb!]
\centering
\includegraphics[width=0.99\textwidth]{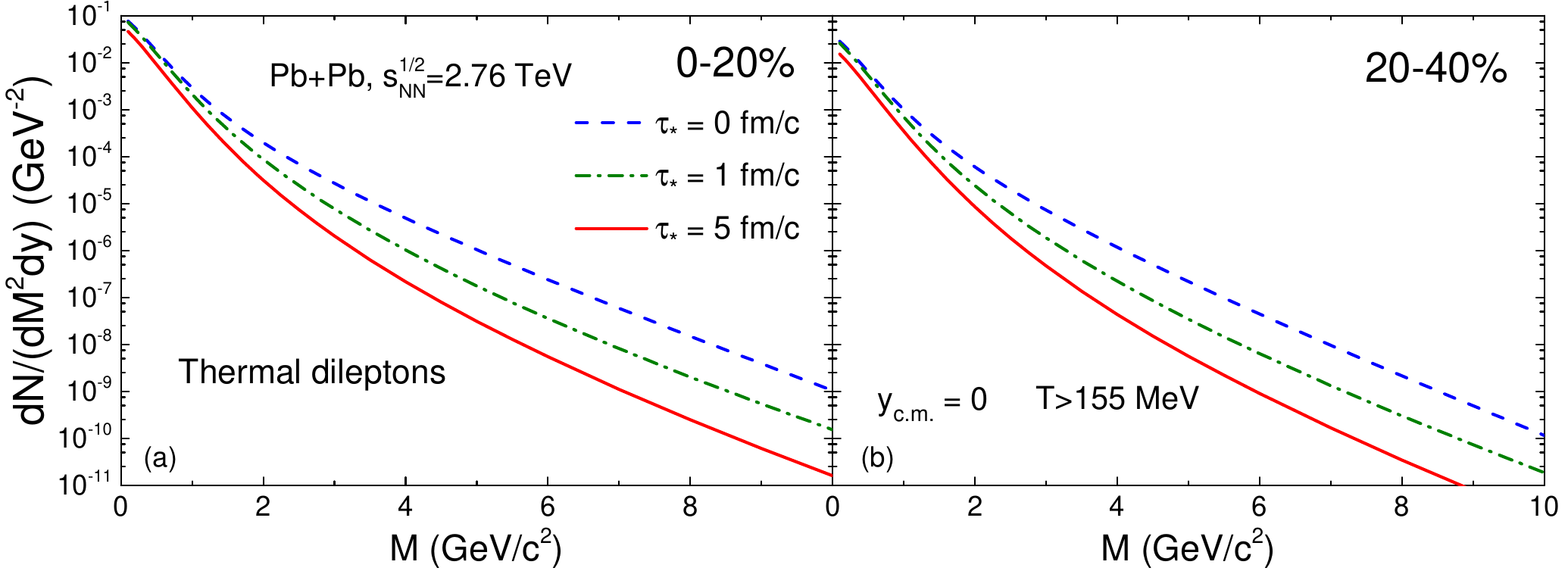}
\caption[]{
Invariant mass distribution of thermal dileptons in {the 0--20\hspm\%~(a)
and 20--40\hspm\%~(b)} central Pb+Pb collisions at $\sqrt{s_{NN}}=2.76~\textrm{TeV}$ calculated
for $\tau_* = 0, 1~\textrm{and}~5~\textrm{fm}/c$\,. {All results correspond to}
the cut-off temperature $T_f=155~\textrm{MeV}$.
}
\label{fig:dilept-dNdM-155}
\end{figure}

We also analyze spectra of thermal dileptons produced in nuclear collisions at the LHC energies. It is assumed that dileptons are produced in the $q\ov{q}\to e^+e^-$ processes.
The additional factor~$\lambda^2$ is introduced into the dilepton production rate which takes into account the quark suppression in a chemically nonequilibrium QGP.
The results of calculating the dilepton mass spectrum in $0-20$\% and $20-40$\%
central Pb+Pb collisions at \mbox{$\sqrt{s_{NN}} = 2.76$~TeV} are shown
in~Fig.~\ref{fig:dilept-dNdM-155}
for the cut-off temperature $T_f=155~\textrm{MeV}$.  One can see that the initial
quark suppression leads to a strong reduction of the dilepton yield at $M\gtrsim 2~\textrm{GeV}$.
Note that we do not include contributions of hard (Drell-Yan) dileptons~\cite{Lin16}  produced in binary collisions of initial nucleons.

\begin{figure}[htb!]
\centering
\includegraphics[width=0.99\textwidth]{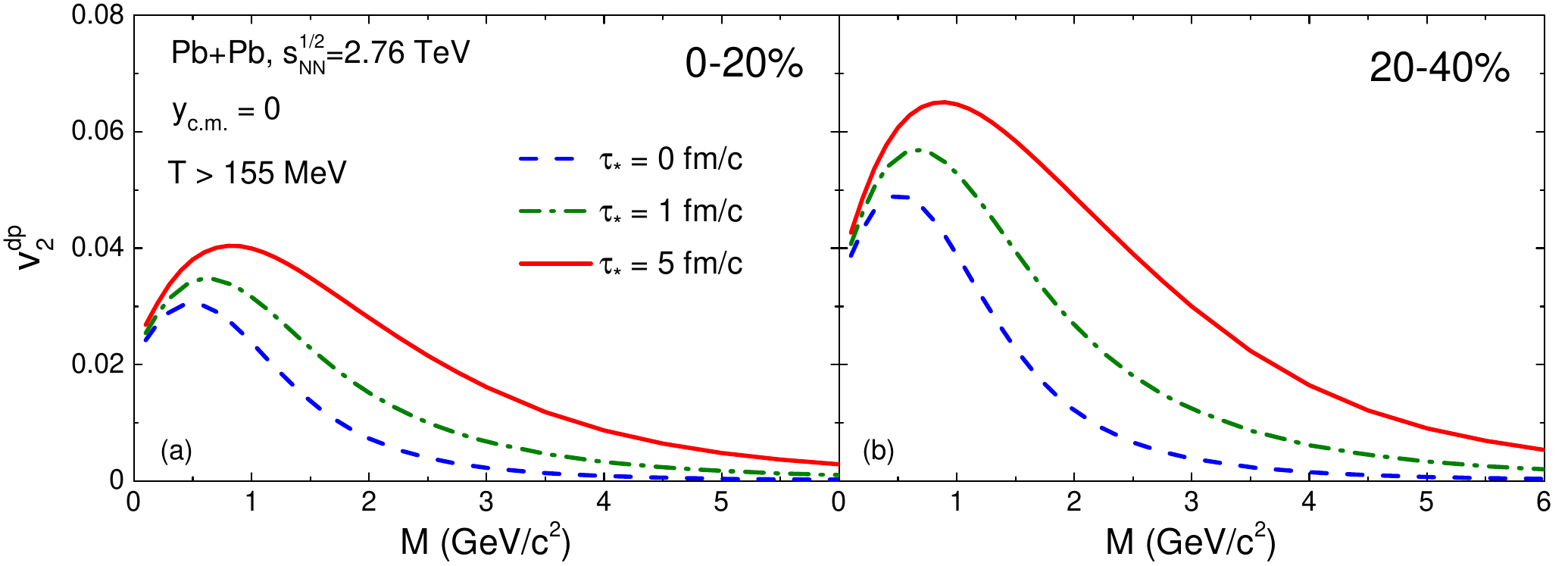}
\caption[]{
Same as Fig.~\ref{fig:dilept-dNdM-155} but for elliptic flow
of thermal dileptons $v_2^{\rm dp}$.
}\label{fig:dilept-v2}
\end{figure}

The elliptic flows of thermal dileptons in the same reaction
are shown in Fig.~\ref{fig:dilept-v2} for several values of $\tau_*$. Similar to direct photons we predict a strong enhancement of the dilepton elliptic flow
as compared to the equilibrium sce\-nario~($\tau_*=0$).

\section{Pure glue scenario at lower energies}
The presented calculations, as well as their comparison to experimental data, show that the pure glue initial scenario does not spoil the existing agreement of hydro simulations with the data at LHC energies. On the other hand, e.g. as seen from Fig.~\ref{fig:T}b, the effect of initial pure gluodynamic evolution, in particular the first-order phase transition at $T\simeq 270$~MeV, is almost completely washed out in most observables.
It is evident that the gluodynamical effects can be more pronounced at lower collision energies where the initial temperature are much closer to the critical temperature $T_c = 270$~MeV of the deconfinement phase transition in the Yang-Mills theory.

In order to investigate this aspect qualitatively, we use the one-dimensional Bjorken model and consider just the two limiting cases: the fully equilibrated (2+1)-flavor evolution and the pure gluodynamic evolution. The corresponding equations of state are shown in Fig.~\ref{fig:latticeEoS}. 
We also consider both the heavy-ion ($A+A$) and the smaller $p+p$ systems.
Unlike for LHC energy, we adopt $\tau_0=0.5$~fm/$c$ in the present analysis.
The initial entropy density $s_0$ at a given collision energy is estimated by using the available data on pion multiplicity. In order to estimate the uncertainties of the obtain results the transverse radius $R$ is varied, namely $R=(6-9)$~fm for $A+A$ collisions and $R=(0.6-0.9)$~fm for p+p collisions.

\begin{figure}[ht]
\centering
\includegraphics[width=0.49\textwidth]{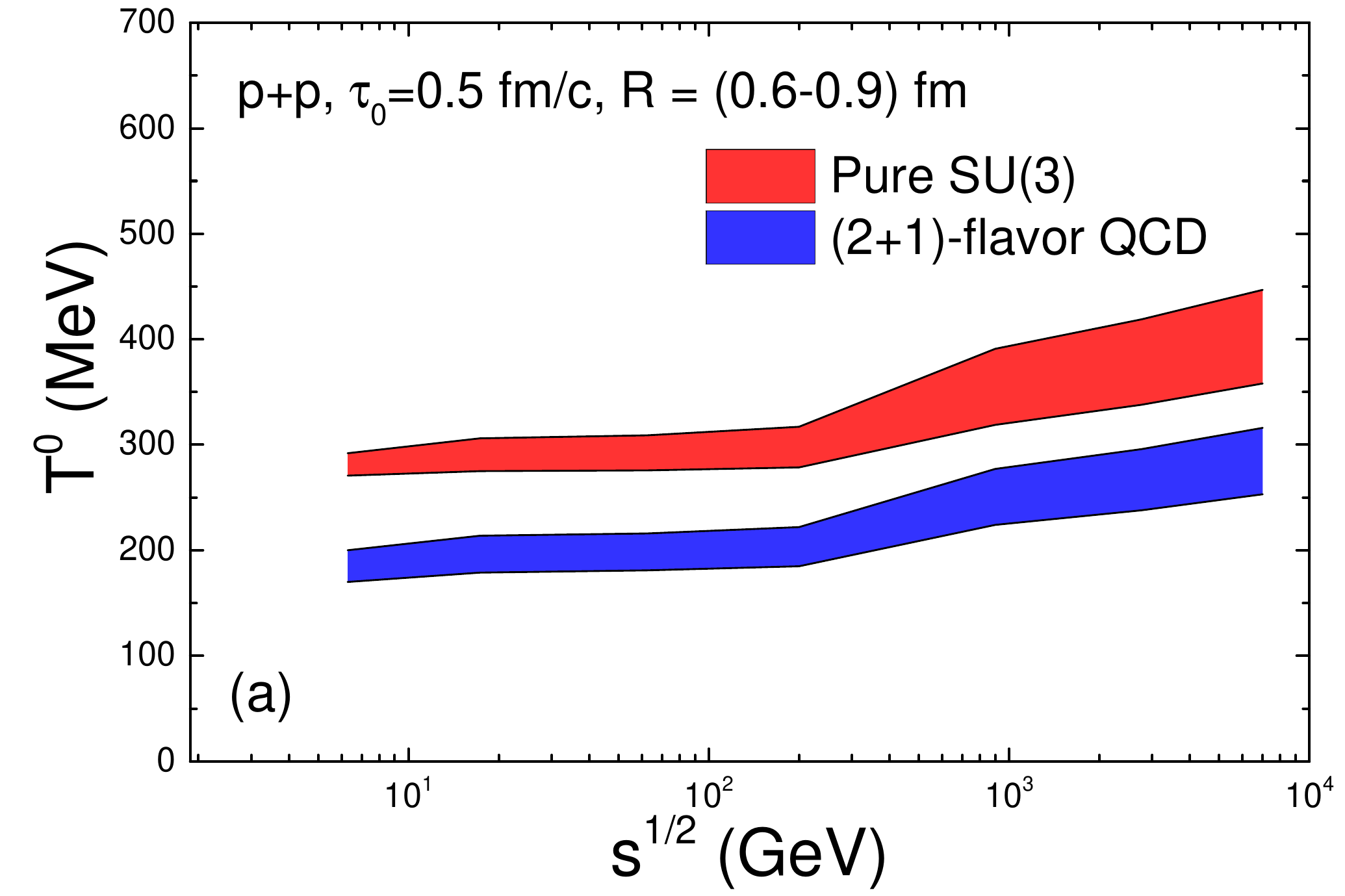}
\includegraphics[width=0.49\textwidth]{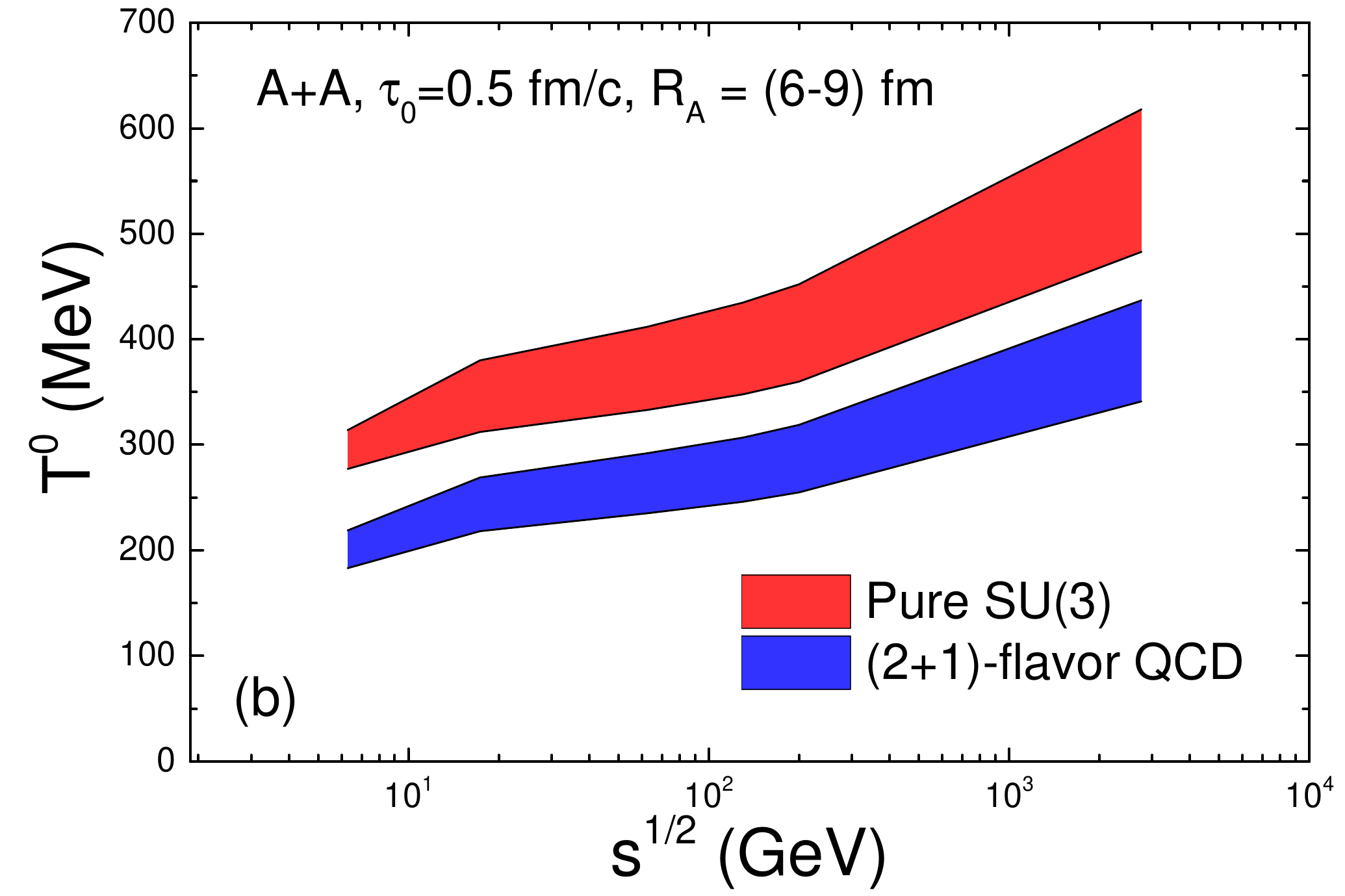}
\caption[]{(Color online)
Dependence of the initial temperature $T^0$
at $\tau_0 = 0.5$~fm/$c$
on the
collision energy for (2+1)-flavor QCD and pure SU(3) scenarios in
(a) $p+p$ and (b) $A+A$ collisions.
The uncertainty bands result from variation of the transverse radius.
}\label{fig:T0-vs-en}
\end{figure}

\begin{figure}[ht]
\centering
\includegraphics[width=0.70\textwidth]{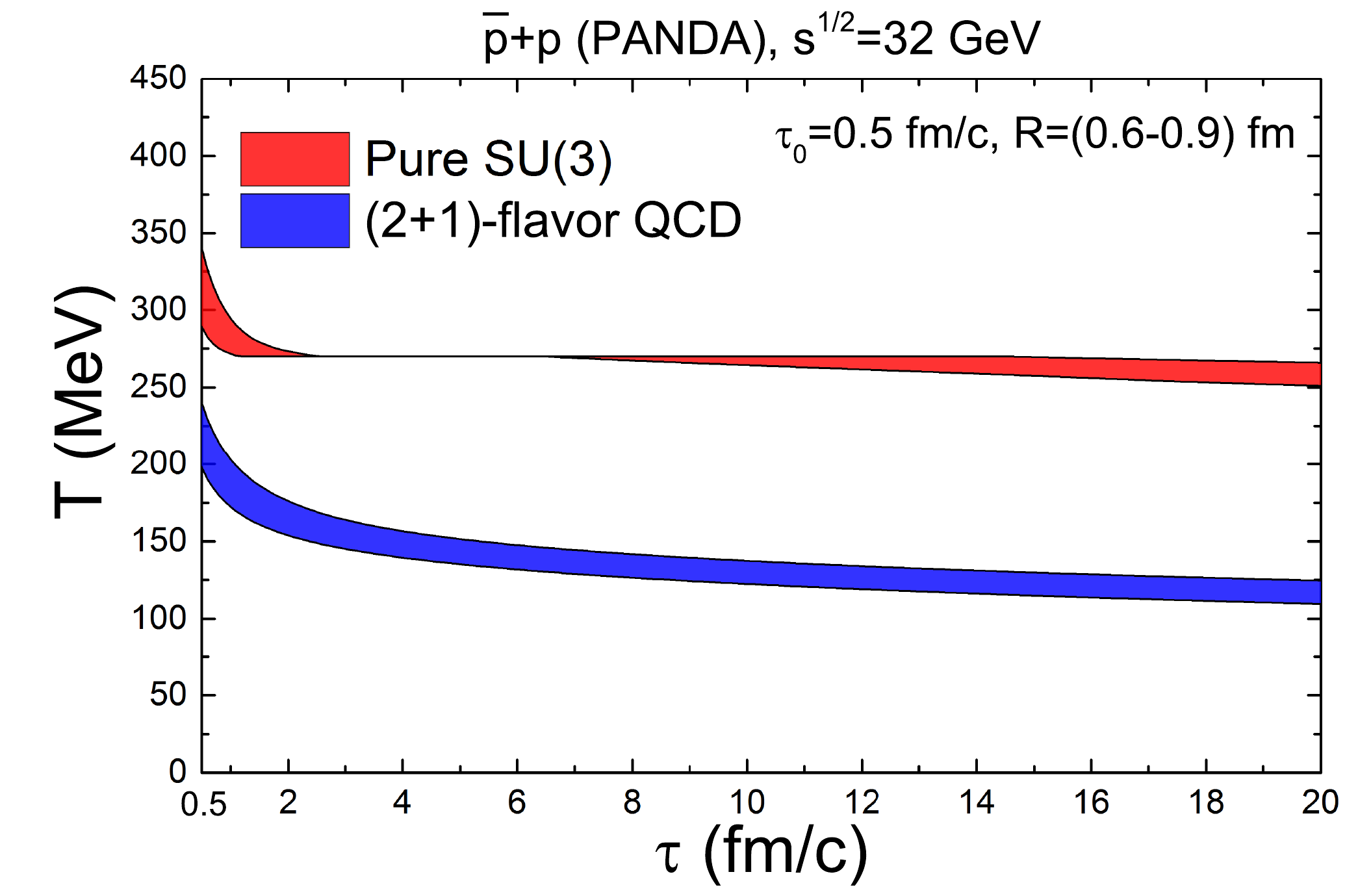}
\caption[]{(Color online)
The $\tau$-dependence of the temperature
for (2+1)-flavor QCD and pure SU(3) scenarios in
$\bar{p}+p$ collisions at $\sqrt{s_{_{\rm NN}}} = 32$~GeV.
The uncertainty bands result from variation of the transverse radius.
}\label{fig:T-vs-tau}
\end{figure}

The values of extracted initial temperature for $p+p$ and $A+A$ are shown in Fig.~\ref{fig:T0-vs-en}. The resulting initial temperature in pure SU(3) case at collision energies $\sqrt{s_{_{\rm NN}}} \lesssim 100$~GeV is rather close to (slightly above) the phase transition temperature of $T_C \simeq 270$~MeV. This especially concerns the p+p collisions.
Thus, these energies look promising for observing the effects of the phase transition in the pure glue scenario.

At smaller collision energies, however, the effect of constituent quarks in colliding nucleons becomes more and more important. In that sense it becomes difficult to consider the matter created in such systems as net-baryon free.
On the other hand, the $\bar{p}-p$ collisions at these energies may be more promising
with regards to the pure glue scenario. The PANDA experiment at FAIR~\cite{PANDA} is,
in principle, capable of running such an experiment in the future, by colliding the proton and (anti)-proton beams at $\sqrt{s_{_{\rm NN}}} \simeq 30$~GeV.
The $\tau$-dependence of the temperature for such a configuration within Bjorken hydrodynamics is depicted in Fig.~\ref{fig:T-vs-tau}. In the pure glue scenario the matter can 
enter the mixed phase at $T=T_c=270$~MeV at the early stage of evolution and
spend a significant amount of time there. In this regard, the future PANDA experiment looks promising in the search of new exotic states of matter.

\section{Summary}
In summary, the aspects and the calculation results of the hydrodynamic modeling of the pure glue initial scenario for hadron and heavy-ion collisions has been presented.
The calculations performed within three different hydro codes
all show a consistent physical picture: the evolution of the pure glue matter in heavy-ion collisions is very different from that of a fully equilibrated QCD matter. The pure glue matter evolves for a much longer time and spends a significant portion of its space-time evolution in the region of mixed phase.
The calculations performed for the LHC energy show that the pure glue initial scenario does not spoil the existing agreement of hydro with the data, in particular with regards to the direct photon yield. The suppression of dilepton yield and the enhancement of its momentum anisotropy may serve as the promising observables to determine the properties of the initial state in Pb+Pb collisions at LHC.
Estimates based on the Bjorken hydrodynamics imply that collisions of smaller systems at smaller collision energies, for instance the $\bar{p}-p$ collisions at the possible PANDA energy of $\sqrt{s_{_{\rm NN}}} \simeq 30$~GeV, are promising for searching new exotic states of matter.

\section*{Acknowledgements}
This work was supported
by HIC for FAIR within the LOEWE program of the
State of Hesse. H.St. acknowledges the support through
the Judah M. Eisenberg Laureatus Chair at Goethe University.
V.V. acknowledges the support from HGS-HIRe for FAIR. The work of M.I.G. was supported by the Goal-Oriented Program of  the National Academy of Sciences of Ukraine and 
the European Organization for Nuclear Research (CERN), Grant CO-1-3-2016.

\end{document}